\documentclass[letterpaper,twocolumn,prl,aps,superscriptaddress,amsmath,amssymb,floatfix]{revtex4-1}
\usepackage{mathptmx}
\usepackage[latin9]{inputenc}
\usepackage{color}
\usepackage{verbatim}
\usepackage{float}
\usepackage{amsmath}
\usepackage{amssymb}
\usepackage{graphicx}
\usepackage{esint}
\usepackage[unicode=true,
 bookmarks=true,bookmarksnumbered=false,bookmarksopen=false,
 breaklinks=false,pdfborder={0 0 1},backref=false,colorlinks=true]
 {hyperref}
\hypersetup{
 linkcolor=magenta,urlcolor=blue,citecolor=blue,pdfstartview={FitH},hyperfootnotes=false}

\makeatletter

\usepackage{textcomp}
\usepackage{epstopdf}

\pdfpageheight\paperheight
\pdfpagewidth\paperwidth

\@ifundefined{textcolor}{}{%
 \definecolor{BLACK}{gray}{0}
 \definecolor{WHITE}{gray}{1}
 \definecolor{RED}{rgb}{1,0,0}
 \definecolor{GREEN}{rgb}{0,1,0}
 \definecolor{BLUE}{rgb}{0,0,1}
 \definecolor{CYAN}{cmyk}{1,0,0,0}
 \definecolor{MAGENTA}{cmyk}{0,1,0,0}
 \definecolor{YELLOW}{cmyk}{0,0,1,0}
}

\usepackage{xcolor}\usepackage{soul}
\definecolor{blue}{rgb}{0,0,1}
\definecolor{red}{rgb}{1,0,0}
\definecolor{green}{rgb}{0,1,0}

\usepackage{soul}

\makeatother

\begin{document}

\affiliation{CAS Key Laboratory of Quantum Information, University of Science and Technology of China, Hefei, Anhui 230026, P. R. China.}
\affiliation{School of Physics and Optoelectronics, South China University of Technology, Guangzhou 510460, P. R. China.}
\affiliation{Interdisciplinary Center for Quantum Information, State Key Laboratory of Modern Optical Instrumentation, College of Optical Science and Engineering, Zhejiang University, Hangzhou 310027, China.}
\affiliation{CAS Center For Excellence in Quantum Information and Quantum Physics, University of Science and Technology of China, Hefei, Anhui 230026,
P. R. China.}
\affiliation{Intelligent Optics $\&$ Photonics Research Center, Jiaxing Research Institute Zhejiang University, Jiaxing 314000, China.}

\title{Nonlinear optical radiation of a lithium niobate microcavity}

\author{Yuan-Hao~Yang}
\thanks{These two authors contributed equally to this work.}
\affiliation{CAS Key Laboratory of Quantum Information, University of Science and Technology of China, Hefei, Anhui 230026, P. R. China.}
\affiliation{CAS Center For Excellence in Quantum Information and Quantum Physics,
University of Science and Technology of China, Hefei, Anhui 230026,
P. R. China.}

\author{Xin-Biao~Xu}
\thanks{These two authors contributed equally to this work.}
\affiliation{CAS Key Laboratory of Quantum Information, University of Science and Technology of China, Hefei, Anhui 230026, P. R. China.}
\affiliation{CAS Center For Excellence in Quantum Information and Quantum Physics,
University of Science and Technology of China, Hefei, Anhui 230026,
P. R. China.}

\author{Jia-Qi~Wang}
\affiliation{CAS Key Laboratory of Quantum Information, University of Science and Technology of China, Hefei, Anhui 230026, P. R. China.}
\affiliation{CAS Center For Excellence in Quantum Information and Quantum Physics,
University of Science and Technology of China, Hefei, Anhui 230026,
P. R. China.}

\author{Mai~Zhang}
\affiliation{CAS Key Laboratory of Quantum Information, University of Science and Technology of China, Hefei, Anhui 230026, P. R. China.}
\affiliation{CAS Center For Excellence in Quantum Information and Quantum Physics,
University of Science and Technology of China, Hefei, Anhui 230026,
P. R. China.}

\author{Ming~Li}
\email{lmwin@ustc.edu.cn}
\affiliation{CAS Key Laboratory of Quantum Information, University of Science and Technology of China, Hefei, Anhui 230026, P. R. China.}
\affiliation{CAS Center For Excellence in Quantum Information and Quantum Physics,
University of Science and Technology of China, Hefei, Anhui 230026,
P. R. China.}

\author{Zheng-Xu~Zhu}
\affiliation{CAS Key Laboratory of Quantum Information, University of Science and Technology of China, Hefei, Anhui 230026, P. R. China.}
\affiliation{CAS Center For Excellence in Quantum Information and Quantum Physics,
University of Science and Technology of China, Hefei, Anhui 230026,
P. R. China.}

\author{Zhu-Bo~Wang}
\affiliation{CAS Key Laboratory of Quantum Information, University of Science and Technology of China, Hefei, Anhui 230026, P. R. China.}
\affiliation{CAS Center For Excellence in Quantum Information and Quantum Physics,
University of Science and Technology of China, Hefei, Anhui 230026,
P. R. China.}

\author{Chun-Hua~Dong}
\affiliation{CAS Key Laboratory of Quantum Information, University of Science and Technology of China, Hefei, Anhui 230026, P. R. China.}
\affiliation{CAS Center For Excellence in Quantum Information and Quantum Physics,
University of Science and Technology of China, Hefei, Anhui 230026,
P. R. China.}

\author{Wei~Fang}
\affiliation{Interdisciplinary Center for Quantum Information, State Key Laboratory of Modern Optical Instrumentation, College of Optical Science and Engineering, Zhejiang University, Hangzhou 310027, China.}
\affiliation{Intelligent Optics $\&$ Photonics Research Center, Jiaxing Research Institute Zhejiang University, Jiaxing 314000, China.}

\author{Huakang~Yu}
\email{hkyu@scut.edu.cn}
\affiliation{School of Physics and Optoelectronics, South China University of Technology, Guangzhou 510460, P. R. China.}

\author{Zhiyuan~Li}
\affiliation{School of Physics and Optoelectronics, South China University of Technology, Guangzhou 510460, P. R. China.}

\author{Guang-Can~Guo}
\affiliation{CAS Key Laboratory of Quantum Information, University of Science and Technology of China, Hefei, Anhui 230026, P. R. China.}
\affiliation{CAS Center For Excellence in Quantum Information and Quantum Physics,
University of Science and Technology of China, Hefei, Anhui 230026,
P. R. China.}

\author{Chang-Ling~Zou}
\email{clzou321@ustc.edu.cn}
\affiliation{CAS Key Laboratory of Quantum Information, University of Science and Technology of China, Hefei, Anhui 230026, P. R. China.}
\affiliation{CAS Center For Excellence in Quantum Information and Quantum Physics,
University of Science and Technology of China, Hefei, Anhui 230026,
P. R. China.}

\date{\today}
\begin{abstract}
The nonlinear optical radiation of an integrated lithium niobate microcavity is demonstrated, which has been neglected in previous studies of nonlinear photonic devices. We find that the nonlinear coupling between confined optical modes on the chip and continuum modes in free space can be greatly enhanced on the platform of  integrated microcavity, with feasible relaxation of the phase-matching condition. With an infrared pump laser, we observe the vertical radiation of second-harmonic wave at the visible band, which indicates a robust phase-matching-free chip-to-free-space frequency converter and also unveils an extra energy dissipation channel for integrated devices. Such an unexpected coherent nonlinear interaction between the free-space beam and the confined mode is also validated by the different frequency generation. Furthermore, based on the phase-matching-free nature of the nonlinear radiation, we build an integrated atomic gas sensor to characterize Rb isotopes with a single telecom laser. The unveiled mechanism of nonlinear optical radiation is universal for all dielectric photonic integrated devices, and provides a simple and robust chip-to-free-space as well as visible-to-telecom interface.
\end{abstract}

\maketitle
\emph{Introduction.-} Cavity-enhanced nonlinear photonic devices offer a compact solution for efficient frequency conversion under low pump power owing to the small mode volume and high quality-factor~\cite{Shen2020,Strekalov2016,Kippenberg2018,Bruch2021,xiao2020ultra}. In particular, the chip-integrated optical microcavity, which benefits from the high-flexibility in dispersion engineering, enables significant nonlinear optics effects among multiple resonant modes with distinct wavelengths by $\chi^{(2)}$ and $\chi^{(3)}$ processes~\cite{Bruch2019,DelHaye2007}, as well as Raman and Brillouin scattering~\cite{Takahashi2013,Eggleton2019,Safavi-Naeini:19}. During last decades, the preparation and micro/nano-fabrication techniques of thin-film optical materials with high nonlinear susceptibility, such as aluminum nitride~\cite{Xiong2012,Jung2013,Soltani2016,Guo2016,wang2021}, gallium nitride~\cite{Bruch2015,Mohamed2017,Stassen2019,Zheng2022}, gallium arsenide~\cite{Kuo2014,Midolo2015,Dietrich2016}, and lithium niobate (LN)~\cite{Lin2019,Pan:19,chengya,liu2021,qi2020,zhou2021chip,xue2021ultrabright,lin2020advances,yuan2021strongly,Liu2020} have been well developed, making integrated microcavities with quality (Q) factor higher than $10^{7}$ obtainable. As a result, an unprecedented conversion efficiency $5\times 10^{4}\ \mathrm{W^{-1}}$ of second-harmonic generation (SHG) and ultra-low-threshold optical parametric oscillators are demonstrated in thin-film LN microcavities~\cite{Lu2020,Lu2021}.

Despite the excellent performance of on-chip nonlinear optics devices, the experimental studies are restricted to the guided modes within the integrated photonic circuits. From the fundamental light-matter interaction point of view, the enhanced nonlinear optical processes should not be limited to the in-plane guided modes. It is anticipated that the potential coupling from the on-chip modes to free-space continuum modes would lead to additional optical radiation channels~\cite{LiTheory2022}, which might impose ultimate performance limitations of integrated devices. For practical applications, the in-plane nonlinear frequency conversion has several drawbacks: Firstly, special geometry designs are required for matching the frequencies and  momentums of target optical modes with very different frequencies~\cite{Guo2016,wang2021,Lu2019}. Secondly, the collection and routing of the converted light for specific optical applications, such as the spectroscopy of gas sample in free space~\cite{Hummon:18,Niffenegger2020}, demands stable and specially designed chip-to-free-space linear mode converters, such as grating, holographic, and meta-surface structures~\cite{lee2007universal,jiang2016whispering,Kim2018,Calafiore2014,Hu2020}. Moreover, these devices usually have a limited working bandwidth.

In this Letter, the nonlinear optical radiation of an integrated LN microring due to the second-harmonic and sum-frequency emissions are experimentally observed. By simultaneously exciting counter-propagating resonant modes with 17.6~mW telecom pump laser, the radiation to vertical direction of 6.57~pW is collected. Compared to previous doubly-resonant SHG process where strict structure design is needed for the phase-matching conditions, the second-harmonic radiation is mode- and pump wavelength-independent. Furthermore, the broadband second-harmonic radiation is applied to the spectroscopy studies of visible transition lines of rubidium atoms by an telecom probe laser, which manifests an unique chip-to-free-space and also visible-to-telecom interface. Our work reveals the unexplored regime of cavity nonlinear photonics, merges the gap between integrated photonics devices and free-space applications.

\noindent 
\begin{figure}[h]
\begin{centering}
\includegraphics[width=\columnwidth]{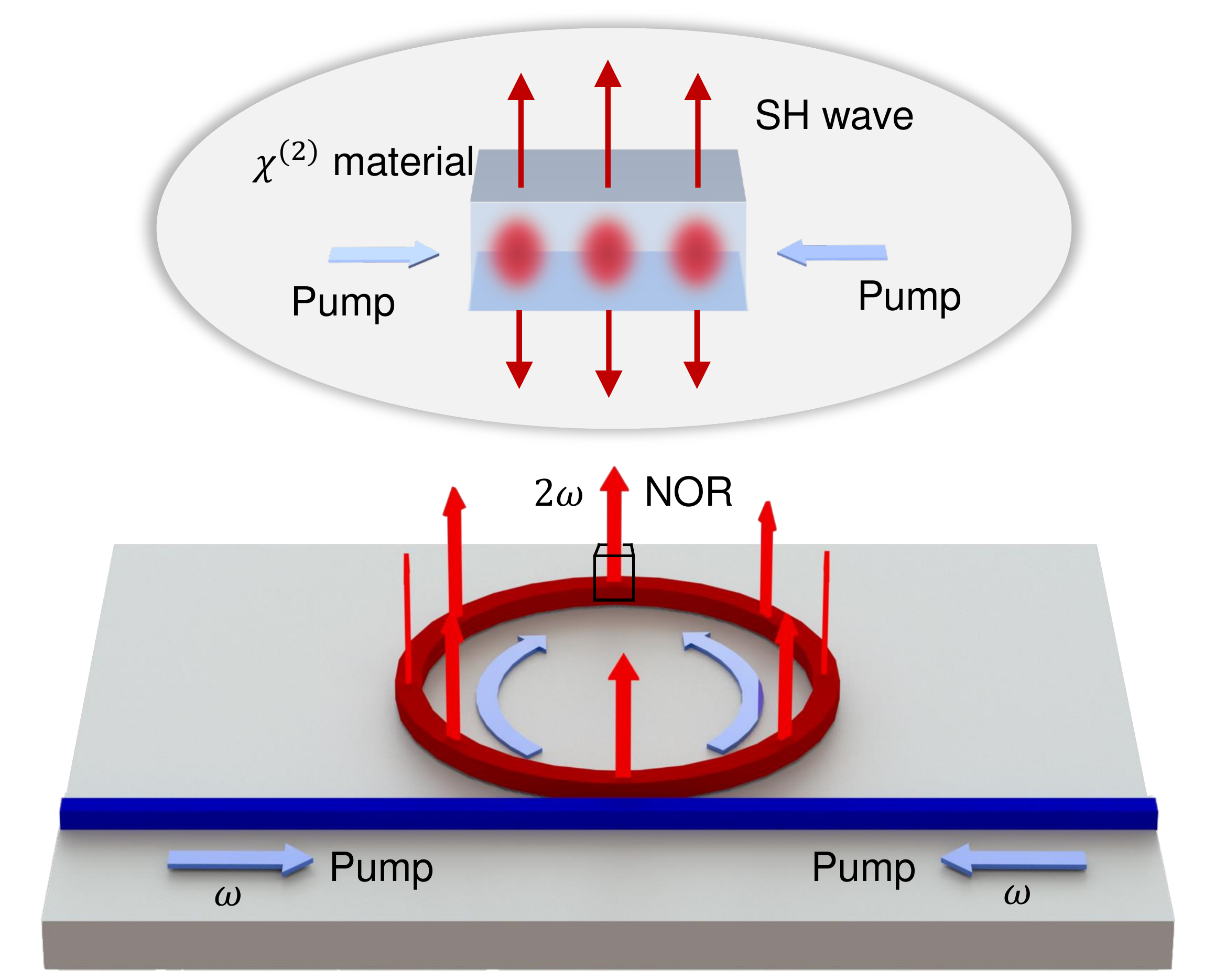}
\par\end{centering}
\caption{Schematic illustration of the nonlinear optical radiation (NOR) process. The blue arrows indicate counter-propagating telecom pump lasers (frequency $\omega$) while the red arrows indicate vertically radiated second-harmonice light (frequency $2\omega$). The inset shows the detail of the nonlinear interaction between guided modes and free-space continuum modes at the cross-section of the microring.}
\label{Fig1}
\end{figure}

 Figure~\ref{Fig1} schematically illustrates the nonlinear optical radiation (NOR) from an on-chip microring cavity through second-harmonic process. Two telecom pump lasers (the blue arrows) with the same frequency ($\omega$) and polarization are coupled into the microring cavity from opposite directions, which excite the degenerate clockwise (CW)  and counter-clockwise (CCW) modes, respectively. As the inset shows, along the microring the standing-wave pump field creates the second-harmonic (SH) nonlinear polarization through the $\chi^{(2)}$ susceptibility of the dielectric material, which is equivalent to nanoparticles oscillating at a frequency of $2\omega$ and radiating light into the free-space~\cite{yu2014single}. Therefore, by selecting an appropriate crystal axis and tuning the polarization of the pump laser to match the nonlinear susceptibility tensor, the SH radiation wave can be collected from the direction perpendicular to the chip. Reversely, an input light from free-space can also couple with the guided modes in the microring by exciting a localized spot of the nonlinear polarizability. It is also anticipated that radiation due to the sum-frequency generation also exists~\cite{LiTheory2022} when selectively driving the counter-propagating pump modes with different frequencies. Comparing to the conventional SH generation between on-chip guided modes, where special geometry design~\cite{Surya2019,Luo2019,briggs2021simultaneous} or periodically poling~\cite{QPM1992,Wang:18,Jankowski2020,Lu2020,hao2020second,Chen2021fan} is required for strict phase-matching or quasi-phase-matching conditions, the NOR is expected to be ultra-broadband and free from the dispersion engineering of the microcavities.

\begin{figure*}[t]
\begin{centering}
\includegraphics[width=0.9\linewidth]{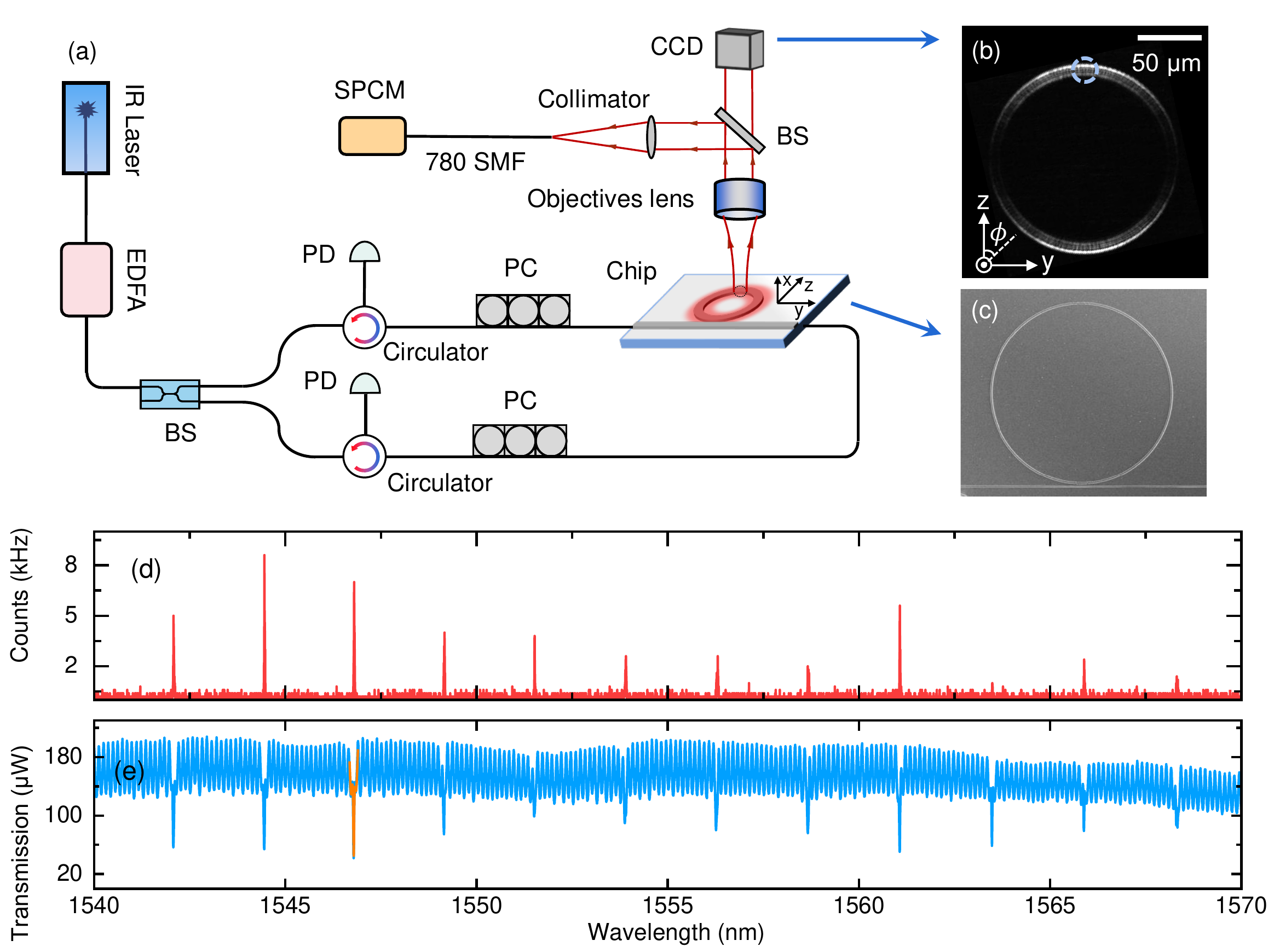}
\par\end{centering}
\caption{(a) Experimental setup. The pump laser is split by a 50:50 BS and launched to the chip from two directions. The radiated light is collected by an objective lens with N.A. of 0.47. IR laser, infrared laser; EDFA, erbium doped fiber amplifier; BS, beam splitter; PC, polarization controller;  PD, photo-detector; SPCM, single photon counting module; 780 SMF, 780~nm single mode fiber. The NOR signal is coupled into a standard 780~nm  single mode fiber which can naturally filter the scattering of telecom pump. The axes on the chip denote crystal orientations of the thin-film lithium niobate, which is corresponding to the axes in (b). (b) The CCD (Basler acA1920-155um) image of the microring when the bi-directional pumps are on-resonance with the cavity, showing the SH radiation. (c) The SEM picture of the device. (d) and (e) show the detected NOR signal and transmission spectrum of pump respectively. The orange line in (e) is fitting of cavity resonance with $Q_{\text{load}}=5.15\times10^4$.}
\label{Fig2}
\end{figure*}

\emph{Experiment setup.-}
The NOR mechanism is investigated by an integrated LN microcavity with the experimental setup shown in Fig.~\ref{Fig2}(a). 
The LN microring is fabricated with 500$\:$nm thick X-cut thin-film LN on a sapphire substrate (NanoLN), and Fig.~\ref{Fig2}(c) shows the SEM image of the waveguide-coupled microring device. The device patterns are defined by hydrogen silses-quioxane (HSQ) resist via electron-beam lithography, and then transferred to LN through the optimized $\mathrm{Ar^{+}}$ etching process by inductively coupled plasma reactive ion etching (ICP RIE) tool. A thickness of 220$\:$nm of LN is etched. The radius of the microring is 70$\:\mu$m, and the widths of the coupling waveguide and microring cavity are $1.72\,\mathrm{\mu m}$ and $1.80\,\mathrm{\mu m}$, respectively. The coupling gap between the waveguide and microring is about $600\,\mathrm{nm}$. Figure~\ref{Fig2}(e) presents the transmission spectrum of the device, with the loaded and intrinsic quality factors of a typical resonance around 1550\,nm [the orange dips in Fig.~\ref{Fig2}(e)] are $0.52\times10^{5}$ and $1.0\times10^{5}$, respectively.

\begin{figure}[t]
\begin{centering}
\includegraphics[width=\linewidth]{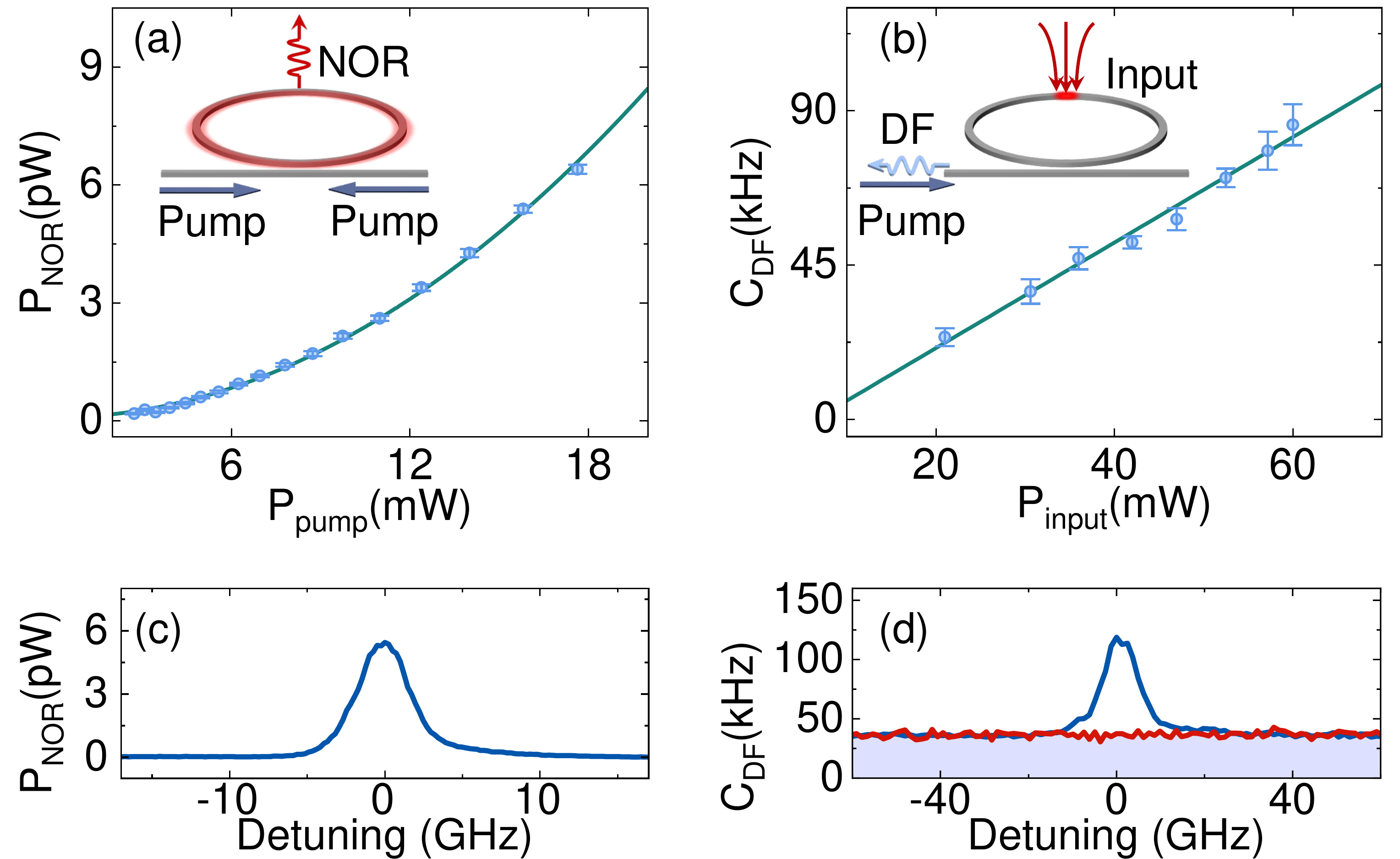}
\par\end{centering}
\caption{Relationship between powers of pump and collected signals. (a), (b) The input-output power dependency of NOR and DF generation, respectively. The dots show the experimental results while the solid lines represent quadratic and linear fitting results. The insets of (a) and (b) are illustrations of the input pump and output signals for NOR and DF generation, respectively. (c) The power of NOR signal when fixing $P_{\text{pump}}=15.8~\mathrm{mW}$ and sweeping pump frequency across cavity resonance. (d) Detected signals DF signals for visible input ($P_{\text{input}}=60.0~\mathrm{mW}$) focused on microring cavity (blue line) and on waveguide (red line), respectively. The purple area denotes the background noise level after filtering while visible input is not injected.}
\label{Fig3}
\end{figure}

A continuous-wave telecom laser is amplified by an erbium-doped fiber amplifier (EDFA) and divided into two paths by a 50:50 fiber beam-splitter. Consequently, the chip is driven by the lasers in two opposite directions using the fiber lens, with an estimated single-side coupling efficiency of $27.4\%$.
Two fiber circulators are introduced in each path for separating and detecting transmitted and reflected signals. Fiber polarization controllers are adjusted for maximum coupling efficiencies to the TE modes of the integrated device. The free-space radiation of the microcavity at visible wavelength ($780\,\mathrm{nm}$) is collected by an objective lens with a $\text{N.A.}=0.47$ and a working distance around 2$\,\mathrm{cm}$. We use a 90:10 beam-splitter to separate the free-space signal into two parts, with 10\% of the signal sending to the CCD for real-time imaging, while the rest (90\%) of the signal is directed to a single mode fiber through a collimator and then fed into a single photon counting module (SPCM).

\textit{NOR.-} When scanning the pump lasers, we observed the periodic shining of the microring from the CCD which only responds to the visible wavelengths. Shown in Fig.~\ref{Fig2}(b) is the image of the device obtained from the CCD, which is lit obviously only when the pump laser is on-resonance with the microring, while the waveguide is invisible. Besides, the radiated visible photons from the microring, which are collected from a small section of waveguide in the microring as the blue circle  shown in Fig.~\ref{Fig2}(b), are counted by SPCM when the wavelength of the telecom laser scans from 1540\,nm to 1570\,nm with a scanning speed of $0.5\,\mathrm{nm/s}$. The resulting spectra of the counts and the transmitted telecom pump laser are recorded [Figs.~\ref{Fig2}(d) and (e)], which show aligned peaks of the visible photons and the dips of the pump laser. The spectra confirm that the radiation occurs for all microring modes when the pump laser is on-resonance with them, and indicate the resonance-enhanced nonlinear frequency up-conversion from telecom to visible wavelengths.

The visible radiation from the chip might be induced by the scattering of the unexpected SH generation between guided modes in the microring and waveguide, however, the scattering process can not explain the following phenomena: (i) The radiation is broadband over a bandwidth exceeding $30\,nm$. (ii) The radiation by the microring is not uniform along the azimuthal angle $\phi$. The large bandwidth indicates the NOR is free of strict phase-matching condition, in contrast to the narrow bandwidth of SH generation between on-chip guided modes. The non-uniform radiation from the microring is attributed to the dependence of the radiation intensity on the $\chi^{(2)}$ nonlinear susceptibility tensor. To be specific, for the X-cut LN and TE microring modes used in this work, the radiation intensity is symmetrical about the Y-crystal axis [labelled in Fig.~\ref{Fig2}(b)] as $\propto \cos^4\phi$ and is the brightest in the upper and lower parts where the largest second-order nonlinear coefficients ($d_{33}=-27\,\mathrm{pm}/\mathrm{V}$) could be accessed. These results agree with our theoretical predictions of NOR~\cite{LiTheory2022}.

To further verify the physical mechanism of the NOR, we study the power-dependent relationship between the input pump and visible radiation.
As shown in Fig.~\ref{Fig3}(a), the power of visible radiation signal increases near quadratically with the pump power. Each data point in Fig.~\ref{Fig3}(a) represents the fitted peak value of the collected radiation spectra for a given pump power. One of the typical spectra of the NOR is shown in Fig.~\ref{Fig3}(c), with the fixing pump power of 15.8~mW. By fitting the experimental data with $P_{\mathrm{NOR}}=A+B\times P_{\mathrm{pump}}^2$, we get $A=0.0879\pm0.0274\,\mathrm{pW}$ and $B=0.0209\pm2.21\times10^{-4}\,\mathrm{pW}\cdot\mathrm{mW}^{-2}$ with high coefficient of determination $R^{2}=0.998$, which agrees with the expected SH generation process. 

\textit{Reversal of NOR.-} According to the intrinsic coherent nonlinear optical effect, the theory of NOR also predicts the different-frequency (DF) conversion from the free-space photons to on-chip guided photons, as a reversal process of the sum-frequency generation in $\chi^{(2)}$ dielectric material. Therefore, we design and carry out the experiment by injecting visible laser to the microring, as schematically illustrated by the inset of Fig.~\ref{Fig3}(b). The DF conversion is realized by reversing the setup of the collection of visible radiation in Fig.~\ref{Fig2}(b), and a vertically input visible photon could be converted to two counter-propagating telecom photons in the microring, with the conversion stimulated by a CCW pump laser. To avoid the noisy background of the generated CW signal due to the backscattering of CCW pump in the micoring, we investigate the non-degenerate case of the NOR by employing microring modes with different mode orders.

Due to the energy conservation condition, the frequency of the generated DF signal is  $\omega_{\mathrm{DF}}=\omega_{\mathrm{input}}-\omega_{\mathrm{pump}}$, so the process is only efficient when $\omega_{\mathrm{DF}}$ meets a resonant mode. Denote the free spectrum range (FSR) of the microring cavity as $\Omega$,
thus we would expect an efficient DF process for $\Delta=\omega_{\mathrm{input}}-2\omega_{\mathrm{pump}}$ being integer multiples of $\Omega$ up to a small cavity dispersion, with the corresponding integer being the mode order with respect to the telecom pump.
For our device, $\Delta\approx600\,\mathrm{GHz}$ is large enough for high performance filtering, corresponding to two FSRs of the cavity. We cascade four band-pass filters to achieve high signal-to-noise ratio (SNR), with the signal and background noise showing in Fig.~\ref{Fig3}(d). For the visible input laser, it was generated by a PPLN frequency doubler pumped by a continuous-wave telecom laser. A half wave plate is applied to control the polarization of visible pump, which can tune the nonlinear interaction to match the strongest nonlinear susceptibility $d_{33}$ by optimizing DF signal.

Figure~\ref{Fig3}(b) shows the linear relationship between the powers of non-degenerate DF signal and the visible input while fixing both the power and frequency of telecom pump on resonance, manifesting the characterization of DF conversion. Figure~\ref{Fig3}(d) shows the DF signals obtained by fixing the on-chip pump power at 10.5~mW, and sweep the detuning of visible input (60.0~mW and 0~mW) with the speed of 1~nm/s. Each data point in Figure~\ref{Fig3}(b) represents the highest value of the peak minus the average of background noise when there is no visible input. The coherent frequency conversion between the free-space light and the on-chip guided modes not only validates the mechanism of NOR, but also promises potential applications for detecting free-space radiations.

\begin{figure}[t]
\begin{centering}
\includegraphics[width=\linewidth]{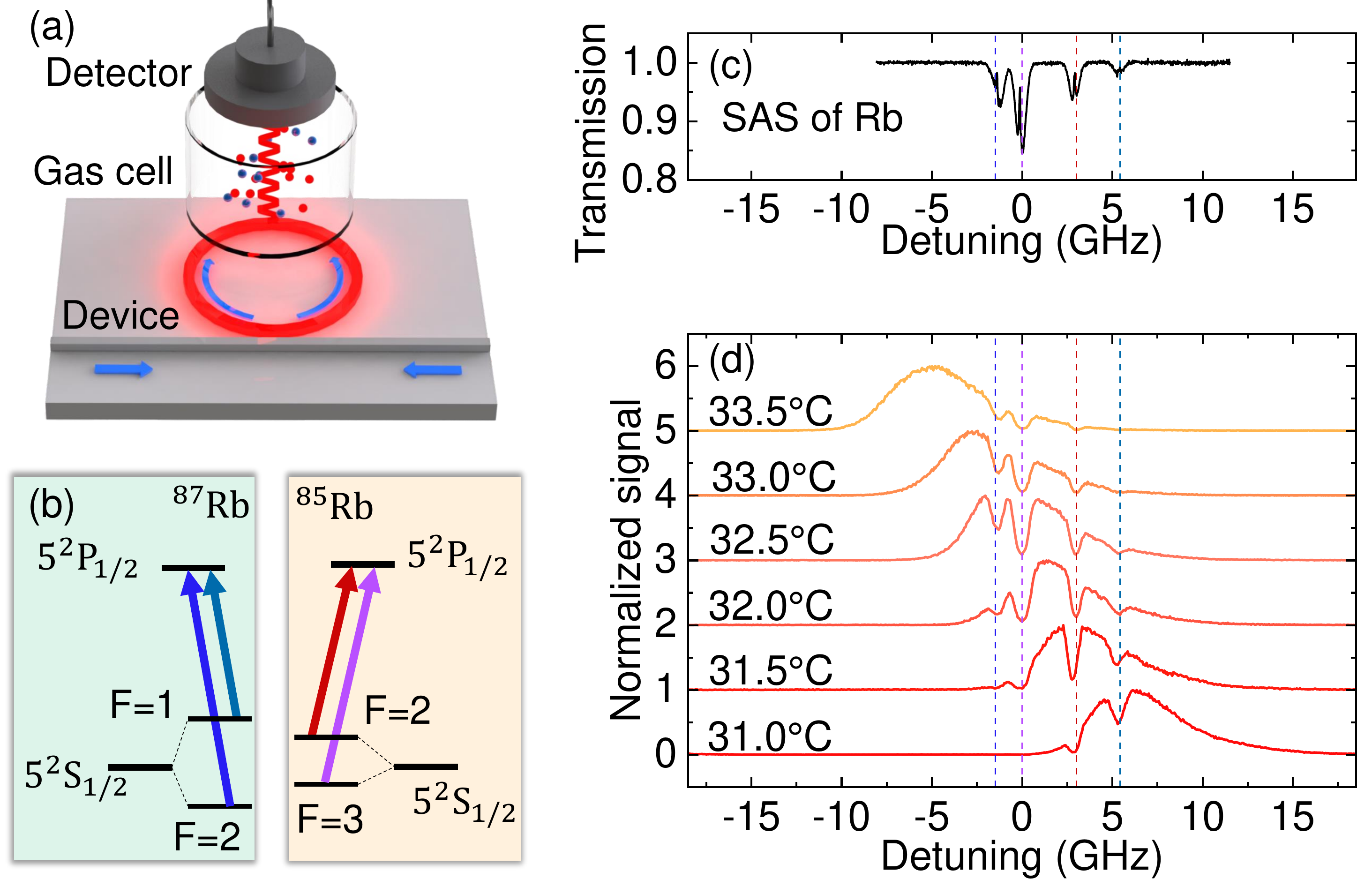}
\par\end{centering}
\caption{(a) Illustration of multi-component gas sensor based on NOR. (b) D2 transitions of $^{\text{85}}\text{Rb}$ and $^{\text{87}}\text{Rb}$ atoms. (c) Saturation absorption spectrum (SAS) of rubidium gas. (d) Tunability of NOR under thermal control of the microcavity. The bandwidth of the device covers all transition lines, shown by the vertical dashed lines.}
\label{Fig4}
\end{figure}

\textit{Application and discussion.-} The demonstrated NOR directly opens up the broadband nonlinear chip-to-free-space conversion channel without specially-designed photonic structures, which offers new opportunities for practical applications of photonic chip. For instance, by leveraging the well-developed on-chip optical soliton microcomb~\cite{Kippenberg2018,Bruch2021}, the NOR could be applied to radiate visible soliton pulses into free space and probe the gas or liquid sample around the chip. Here, we conduct a proof-of-principle application of the NOR for sensing atomic gas with a commercial telecom laser. As shown in Fig.~\ref{Fig4}(a), we insert a cylinder glass cell filling with sample atoms above the chip, and measure the transmitted NOR signal. There are two isotopes of Rubidium atom, i.e. $^{\text{85}}\text{Rb}$ and $^{\text{87}}\text{Rb}$, in the cell which have been extensively investigated in quantum information processing, atomic clock, and precision measurement~\cite{Hummon:18,doi:10.1126/sciadv.aax6230,Gundavarapu2019,Xie2019}. When NOR signal passes through the cell, it would be absorbed by rubidium atoms if its frequency matches the D2 atomic transition lines shown in Fig.~\ref{Fig4}(b).

The broadband nature of NOR makes it possible by simply tuning the frequency of the resonant mode, which otherwise would be difficult for SHG between guided modes and also requires additional chip-fiber or chip-to-free-space coupling apparatus. Here, we use an off-chip heater to finely tune the resonance frequency of the microring cavity to match the transition line at around 780.25~nm. The normalized transmission spectra are shown in Fig.~\ref{Fig4}(d), with four dips corresponding to each two transitions of $^{\text{85}}\text{Rb}$ and $^{\text{87}}\text{Rb}$, which agree with the reference saturated absorption spectrum of Rubidium shown in Fig.~\ref{Fig4}(c). According to the measured spectra, it is possible to distinguish the type of atoms and their concentrations. The tunability and controllability of the system are characterized with the following method: while gradually changing the temperature of the heater, we find that the relative positions of the NOR signal peak shift step by step, which scans across all the absorption dips of Rb atoms. By utilizing the four-wave mixing in the microring cavity for optical frequency comb generation, the NOR can be incorporated to realize high-precision spectroscopy and sensing of various types of gas covering a wide range of wavelengths~\cite{Tan2021,Bao2021}, as well as optical clock by referencing to visible-wavelength atomic transitions~\cite{Hummon:18,doi:10.1126/sciadv.aax6230} based photonic chip. 

It is worth noting that the collected NOR from a focus area with a radius of only about $2\,\mathrm{\mu m}$ is 6.57~pW for only a 17.6~mW on-chip pump power. Considering the radiation of the whole microring and taking account of the loss of our apparatus, the NOR conversion efficiency of our device is estimated to be above $1.50\times10^{-7}$. It is anticipated that the conversion efficiency could be further increased with a higher Q-factor and a higher $P_{\mathrm{pump}}$. For example, with the state-of-art Q-factor of $10^{7}$~\cite{Gao:22} and an on-chip pump power of 100~mW, we predict a conversion efficiency of $4.75\%$. Therefore, the NOR effect could significantly modify the performance of high-quality factor microcavities, and the NOR effect could be more profound for interfacing the free space and photonic chip.

\emph{Conclusion.}- The nonlinear optical radiation of integrated photonic device is experimentally revealed by investigating the free-space second-harmonic radiation of a LN microring resonator. The NOR process as well as its reversal free-space to chip different-frequency conversion process are experimentally verified. The NOR is significant for integrated photonic devices due to the enhanced light-matter interactions,
 and the principle is extensible to other nonlinear optics processes. Thus, we expect similar phenomena could be observed in photonic chips made by other materials. In additional to potential applications for interfacing free-space and photonic chips, the revealed NOR also appeals for more careful investigation and design of photonic devices, especially for quantum devices and precision measurement instruments that NOR might introduce non-negligible and un-avoidable noise photons.

\begin{acknowledgements}
This work was funded by the National Key Research and Development Program (Grant Nos.~2017YFA0304504, 2018YFA0306200), the National Natural Science Foundation of China (Grant Nos.~11874342, 11904316, 11922411, 12104441, 11934012, 91850107, 12174116 and U21A6006), Anhui Provincial Natural Science Foundation (Grant Nos.~2008085QA34 and 2108085MA22), Guangdong Innovative and Entrepreneurial Research Team Program (Grant No. 2016ZT06C594),  Science and Technology Project of Guangdong (2020B010190001). It was also supported by the Fundamental Research Funds for the Central Universities, and the State Key Laboratory of Advanced Optical Communication Systems and Networks, China. This work was partially carried out at the USTC Center for Micro and Nanoscale Research and Fabrication.
\end{acknowledgements}

\smallskip{}

\end{document}